# Electronic Structure of Few-Layer Graphene: Experimental Demonstration of Strong Dependence on Stacking Sequence

Kin Fai Mak[1], Jie Shan[2], and Tony F. Heinz[1*]

[1]Departments of Physics and Electrical Engineering, Columbia University, 538 West 120th St., New York, NY 10027, USA

[2] Department of Physics, Case Western Reserve University, 10900 Euclid Avenue, Cleveland, OH 44106, USA

Abstract

The electronic structure of few-layer graphene (FLG) samples with crystalline order was investigated experimentally by infrared absorption spectroscopy for photon energies ranging from 0.2 – 1 eV. Distinct optical conductivity spectra were observed for different samples having precisely the same number of layers. The different spectra arise from the existence of two stable polytypes of FLG, namely, Bernal (AB) stacking and rhombohedral (ABC) stacking. The observed absorption features, reflecting the underlying symmetry of the two polytypes and the nature of the associated van Hone singularities, were reproduced by explicit calculations within a tight-binding model. The findings demonstrate the pronounced effect of stacking order on the electronic structure of FLG.

*Corresponding author: tony.heinz@columbia.edu

Many of the intriguing properties of graphene [1] are a consequence of its two-dimensional (2D) electronic band structure. Electrons in a single-layer graphene exhibit a characteristic linear dispersion relation between energy and momentum near the K-point of the Brillouin zone [1, 2]. Interlayer coupling in few-layer graphene (FLG) leads to a dramatic change of this electronic structure, with the emergence of hyperbolically dispersing bands [3, 4]. Further, the electronic properties of FLG are predicted to be highly sensitive to crystallographic stacking sequence [4-14]. To date the role of stacking has, however, been overlooked in experimental studies of FLG with crystalline order.

In this Letter we report unambiguous experimental evidence from infrared absorption spectroscopy of the existence of stable FLG of both Bernal (AB) and rhombohedral (ABC) crystal stacking. Our results reveal strikingly different 2D band structure in these two polytypes, reflecting underlying difference in symmetry. In rhombohedral FLG, the lower crystallographic symmetry compared with the usual Bernal stacking leads to energy bands with extrema shifted away from the K-point [15]. This gives rise to 1D-like van Hove singularities in the joint density of states (JDOS) [8] and to prominent optical absorption peaks. These singularities differ from the step features, characteristic of 2D systems [16], that are seen in Bernal-stacked FLG. The existence of stable polytypes of FLG provides a new tool to tailor the electronic structure for both fundamental studies and applications [4-8, 14, 17]. In particular, the effect of externally applied electric or magnetic fields is predicted to be highly sensitive to the stacking sequence [4-8, 12, 17]. A perpendicular electric field is, for example, expected to induce band crossing (and a gapless state) in Bernal stacked FLG, but band repulsion (and an insulating state) in rhombohedrally stacked samples [5, 17]. Using the ready identification of different stacking sequence in FLG by the methods reported here, researchers will now be able to address such predictions experimentally.

We employed the infrared absorption spectroscopy to investigate the low-energy electronic excitations of FLG samples. Optical conductivity spectra spanning a photon energy range of 0.2 – 1 eV were measured. The investigations were performed using the National Synchrotron Light Source at Brookhaven National Laboratory (U2B beamline) to generate the required infrared radiation. Broadband infrared radiation from the synchrotron was focused onto the sample with a 32× reflective objective to achieve a spot size below 10 $\mu m$ in diameter. The radiation reflected from the sample was collected and spectrally analyzed using a Fourier-transform infrared spectrometer (Nicolet system with a HgCdTe detector). For direct interpretation of the infrared reflectance data, we chose the transparent substrate of $SiO_2$ (Chemglass). Graphene samples were deposited by mechanical exfoliation of kish graphite (Toshiba) on substrates cleaned by sonication in methanol. FLG samples of at least several hundreds of $\mu m^2$ in size were selected for the infrared measurements. The thickness of the FLG samples was reliably determined by their reflectance in the visible, since in this spectral range each individual layer in FLG absorbs 2.3% of the light [18, 19].

To determine the optical sheet conductivity, $\sigma(\hbar\omega)$, of the FLG samples as a function of photon energy $\hbar\omega$, we recorded reflectance spectra of both the FLG films on the $SiO_2$ substrate ($R_{FLG/sub}$) and of the bare substrate ($R_{sub}$). For a thin film on a transparent substrate, which accurately describes the situation of the FLG on the $SiO_2$ substrate, we can then obtain the optical conductivity directly from the fractional change of the reflectance as $\delta_R = \frac{R_{FLG/sub} - R_{sub}}{R_{sub}} = \frac{4}{n_{sub}^2 - 1}\frac{4\pi}{c}\sigma$ [18]. Here $c$ denotes the speed of light in vacuum and $n_{sub}$ is the frequency-dependent refractive index of the SiO$_2$ substrate. The result is equivalent, we note, to the determination of the absorbance $A(\hbar\omega)$ of the FLG sample, which is related to the sheet conductivity by $A(\hbar\omega) = [4\pi/c]\sigma(\hbar\omega)$.

We investigated FLG of various thicknesses up to 8 layers. In the following discussion, we will focus on 4-layer graphene samples for a direct comparison to models. We choose this case because the influence of doping and other environmental effects on the infrared conductivity of the tetralayers is expected to be reduced compared to trilayers, while the tetralayer electronic structure still remains relatively simple. Results of trilayer and other FLG samples will be reported in a separate publication.

For the ten tetralayer samples that were studied by infrared absorption spectroscopy, we observed two distinct groups of response in the optical conductivity spectra, with six samples belonging to group 1 and four belonging to group 2. The experimental optical sheet conductivity, $\sigma(\hbar\omega)$, as a function of photon energy $\hbar\omega$ for the two distinct groups of response is presented in Fig. 1a. We note that for both groups, $\sigma(\hbar\omega)$ becomes quite flat at higher photon energies and converges to a value of 4 x $\pi e^2/2h$ (dashed lines), *i.e.*, to 4 times the optical conductivity of graphene monolayer [18, 19]. This is the characteristic signature of tetralayer graphene. For photon energies higher than the interlayer coupling, tetralayer graphene behaves much like four independent graphene monolayers and its optical conductivity is nearly independent of the stacking sequence. On the other hand, the lower-energy part of $\sigma(\hbar\omega)$ for samples in group 1 and 2 is quite distinct: the absorption features in group 2 are more prominent than those in group 1, and the peaks appear at different energies. The main absorption features are observed at 0.24 eV (transition 1) and 0.58 eV (transition 2) for samples in group 1; and at 0.26 eV (transition 1) and 0.35 eV (transition 2) for samples in group 2.

We now consider the role of the crystallographic stacking sequence on the properties of FLG. Hexagonal or AA stacking, where each layer is placed directly on top of another, is the simplest crystallographic structure by stacking individual graphene layers. This arrangement is, however, known to be less stable [20] than the one in which successive layers are displaced along the direction of the honeycomb lattice by a carbon-carbon bond length (Figure 2). Three distinct positions (A, B, and C) of the hexagonal lattice can thus be generated [4, 5, 8, 9, 12]. For bilayers, there is only one distinct stacking structure; for trilayers we can generate either ABA (Bernal) or ABC (rhombohedral) stacking. Considering arbitrary arrangements of these low-energy configurations for adjacent layers, we have $2^{N-2}$ possible stacking sequences for FLG of $N$ layers. Thus, in the case of tetralayer graphene, there are 4 possible stacking sequences: ABAB (Bernal), ABCA (rhombohedral), ABAC, and ABCB. The latter two are equivalent stacking sequences related by inversion symmetry and have been predicted by first-principles electronic structure calculations to be unstable [5]. We therefore expect only two stable polytypes for tetralayer graphene, namely, Bernal (ABAB) and rhombohedral (ABCA) stacked tetralayer. We note that for bulk graphite, Bernal stacking is the thermodynamically stable form, while rhombohedral stacking is a metastable, but known to coexist in small quantities [21].

The electronic states and optical conductivity of Bernal and rhombohedral tetralayer graphene structures can be readily analyzed within the context of a tight-binding (TB) model. To capture the main features observed in the infrared conductivity spectrum, we included in the model only the nearest in-plane and out-of-plane interactions. As in the literature, we describe these, respectively, by the parameters $\gamma_0$ = 3.16 eV and $\gamma_1$ = 0.37 eV [22]. Figure 3 shows the resulting band structure near the K-point for both Bernal and rhombohedral tetralayer graphene samples. The Figure also illustrates the different manifestations of the nearest-neighbor interactions for these two structures, using the schematic representation introduced by Min *et al* [12].

The electronic structure of both Bernal (Fig. 3a) and rhombohedral (Fig. 3b) tetralayer graphene consists of four pairs of conduction and valence bands, but with entirely different topologies near the K-point. First, there are two pairs of low-energy hyperbolic bands (c1, v1, c2, and v2) near the K-point for ABAB stacking, but only one pair of low-energy bands (c1, v1) for ABCA stacking. In the latter case, the bands disperse away from the K-point with momentum as $k^4$, rather than as $k^2$. This difference reflects the hybridization of four isolated (interlayer uncoupled) sites in neighboring layers for ABAB stacking, compared with hybridization of two isolated sites in the top and bottom graphene layers for ABCA stacking [12]. Second, for ABAB stacking there are two pairs of split-off hyperbolic bands (c3, v3, c4, and v4), with split-off energies of $0.60\gamma_1$ and $1.61\gamma_1$ at the K-point [13, 23, 24]. In contrast, there are three pairs of split-off bands for ABCA stacking: the 'wizard hat' shaped (c2, v2) bands, the hyperbolic (c3, v3) bands and the linear (c4, v4) bands. They overlap at the K-point at an energy of $\gamma_1$. One particularly important feature is the energy minimum (or maximum) of the 'wizard hat' bands at $E_{min} \approx 0.30$ eV, which occurs away from the K-point because of the lower crystallographic symmetry of the rhombohedral structure [15].

The different topology of the band structure for the FLG samples of different stacking sequence has direct implications for the infrared absorption spectrum. The first such manifestation concerns the optical selection rules. For light at normal incidence at the basal plane, only transitions between the bands of the same effective masses {v1, v4} →{c1, c4} and {v2, v3}→{c2, c3} are allowed for ABAB stacking [23, 24], while for the lower-symmetry ABCA structure no obvious selection rules apply. Second, the nature of the van Hove singularities at the critical points in the materials' JDOS is entirely different for the two cases (Fig. 3). For ABAB stacking, the band extrema at the K-point produce finite discontinuities at the split-off energies and twice of these energies, followed by a linear increase with energy $E$. These step-singularities are characteristic for 2D systems at the band minimum/maximum [16]. In contrast, for ABCA stacking, in addition to the step singularities, divergences varying as $1/\sqrt{E}$, characteristic of 1D systems, emerge around energies $E_{min}$ and $2E_{min}$. These 1D-like singularities, like those analyzed earlier by Guinea *et al.* [8], arise from bands being parallel along a ring about the K-point of the Brillouin zone where the 'wizard-hat' (c2, v2) bands have their extrema. These divergent singularities are expected to lead to strong and narrow resonance features in the optical conductivity spectrum.

For a detailed comparison of experiment with the TB model, we performed explicit numerical calculations for the optical conductivity spectra for both Bernal and rhombohedral tetralayer graphene using the Kubo formula. In the calculation a phenomenological broadening of 20 meV was included. The samples were assumed to be undoped and at room temperature and were excited by infrared radiation at normal incidence. Simulation (Fig. 1b) is in excellent agreement with experiment (Fig. 1a). It reproduces the main resonance features for each group of response in the optical conductivity spectra, including both the peak positions and conductivity values within a 10% error. For ABAB stacking (group 1) [25] broadened step-function-like absorption features were predicted and observed, respectively, at 0.24 and 0.24 eV for transition 1, and at 0.61 and 0.58 eV for transition 2. Here resonance feature 1 arises from transitions v2→c3 and v3→c2; and feature 2 from transitions v1→c4 and v4→c1. The experimental response for group 2 matches the optical conductivity spectrum calculated for ABCA stacking [25]. In this case, the strongest infrared absorption feature was predicted and observed, respectively, at 0.29 and 0.26 eV. It corresponds to transition 1 (v1→c2, v2→c1) around $E_{min}$, where the JDOS diverges. The second prominent absorption feature was predicted and observed, respectively, at 0.36 and 0.35 eV. It arises from transition 2 (v1→c3, v3→c1) around the split-off energy of $\gamma_1$.

Some limitations of the simple TB model, however, are observable in a more careful comparison with the experimental data. The theory, for example, predicts peak positions up-shifted by as much as 30 meV compared with experiment. Further, the third (weaker) resonance feature expected for rhombohedral graphene tetralayers around 0.67 eV is difficult to identify in the experimental data, presumably because of broadening. Also the line shapes of the experimental features are not fully reproduced by the theory. Some of these discrepancies can be accounted for by the inclusion of additional coupling parameters within the TB model. However, many-body interactions [26], doping, and, perhaps, doping-induced band-gap opening [27-32] may also be of significance. Further theoretical investigations are clearly warranted.

The authors acknowledge support from the National Science Foundation under grant CHE-0117752 at Columbia and grant DMR-0907477 at Case Western Reserve University; from DARPA under the CERA program; and from the New York State Office of Science, Technology, and Academic Research (NYSTAR). The synchrotron studies were supported by the National Synchrotron Light Source at Brookhaven and the Center for Synchrotron Biosciences, Case Western Reserve University under grant P41-EB-01979 with the National Institute for Biomedical Imaging and Bioengineering.

**Figure:**

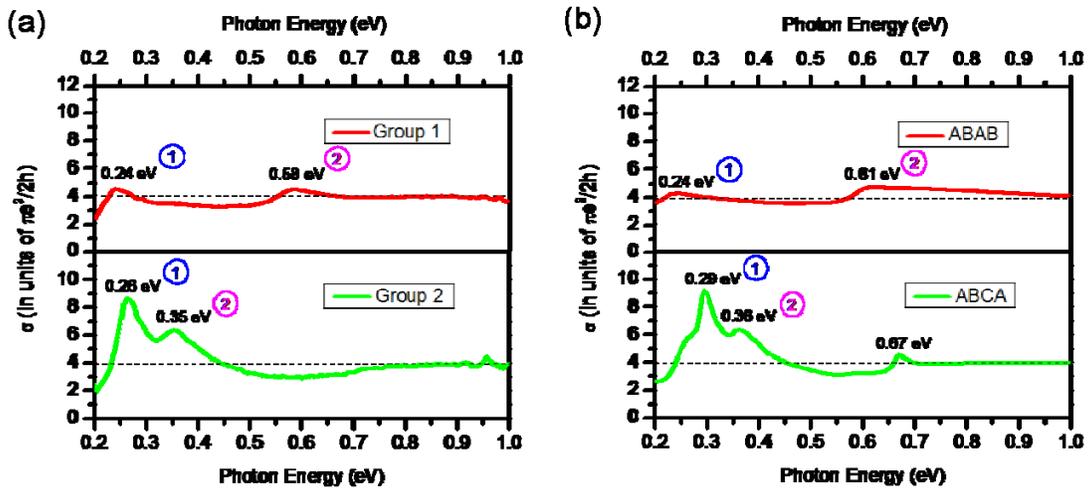

Fig. 1. Infrared optical conductivity spectra of tetralayer graphene samples. (a) Two distinct groups of response in the infrared sheet conductivity spectra $\sigma(\hbar\omega)$ observed in the experiment for tetralayer graphene samples at room temperature. (b) Calculated $\sigma(\hbar\omega)$ for ABAB (Bernal) and ABCA (rhombohedral) stacks by using the Kubo formula and the tight-binding model with the nearest in-plane, $\gamma_0$ (= 3.16 eV), and out-of-plane coupling parameter, $\gamma_1$ (=0.37 eV). In the calculation the samples were assumed to be undoped and at room temperature. A phenomenological broadening of 20 meV was used.

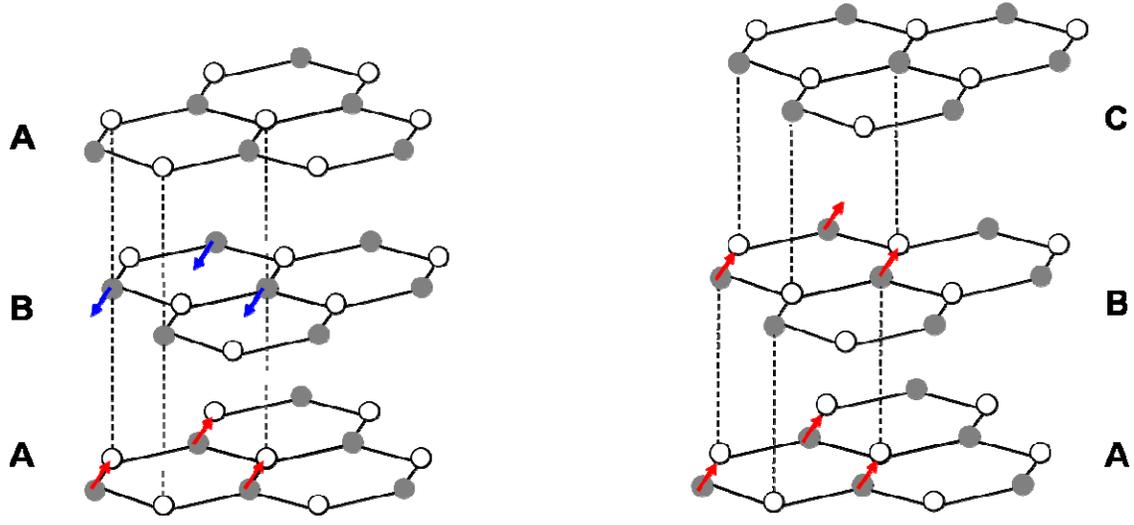

Fig. 2. Crystallographic stacking of few-layer graphene. A, B, and C are three distinct positions of the hexagonal lattice when graphene layers are stacked. They can be generated by displacing atoms of either sublattice $\alpha$ or $\beta$ (filled or empty circles) along the honeycomb lattice by a carbon-carbon bond length (arrows). Left is an ABA (Bernal) structure and right is an ABC (rhomobohedral) structure. The two polytypes can be obtained by displacing each adjacent graphene layer, respectively, in an alternating forward and backward direction (Bernal) and a straight forward direction (rhomobohedral).

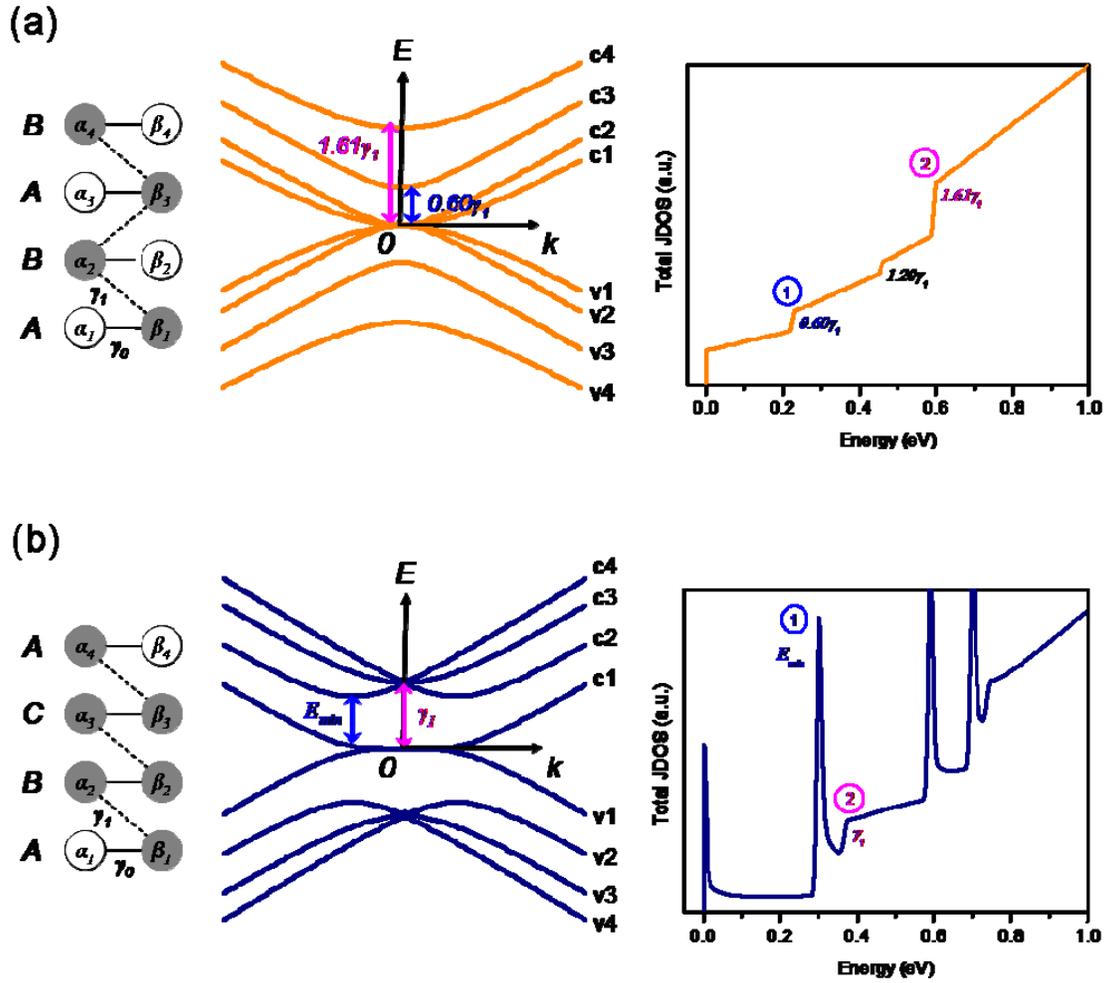

Fig. 3. Electronic structure of tetralayer graphene. (a) ABAB (Bernal) stack. (b) ABCA (rhombohedral) stack. Left, Schematic atomic arrangement: atoms in sublattices $\alpha$ and $\beta$ are represented, respectively, as filled and unfilled circles for interlayer connected and isolated sites; the intra- and inter-layer coupling are denoted by solid and broken lines. Middle, Electronic band structure near the K-point of the Brillouin zone. Right, Total joint-density of states (JDOS) for allowed transitions within the experimentally relevant range of energy.